\documentstyle[epsf]{npbproc}

\begin{document}

\title{On the Dynamics of Light Wilson Quarks}
\author{A. D. Kennedy and Robert G. Edwards\\
        Supercomputer Computations Research Institute\\
        Florida State University\\
        Tallahassee, FL 32306-4052, USA\\
        Internet: adk@scri1.scri.fsu.edu, edwards@mailer.scri.fsu.edu}

\date{}

\runtitle{On the Dynamics of Light Wilson Quarks}
\runauthor{A. D. Kennedy and Robert G. Edwards}

\volume{XXX}
\firstpage{1}
\lastpage{3}

\begin{abstract}
       We describe recent results obtained as part of the High Energy
       Monte Carlo Grand Challenge (HEMCGC) project concerning the
       behaviour of lattice QCD with light dynamical Wilson quarks. We
       show that it is possible to reach regions of parameter space with
       light pions $m_\pi\ll0.2/a$, but that the equilibration time for
       such a system is at least of the order of 1,000 unit-length
       Hybrid Monte Carlo (HMC) trajectories (about a Gigaflop/sec-year).
       If the Hybrid Molecular Dynamics (HMD) algorithm
       is used with the same parameters it gives incorrect results. 
\end{abstract}

\maketitle


\section{Introduction}
%

We set out with the objective of computing the properties of QCD with $m_\pi
\approx 0.25/a$ on a $16^3\times32$ lattice at $\beta=5.3$. This coupling is
probably slightly too strong to be in the scaling region, but our calculation
is intended to be only an early step in the study of QCD with Wilson quarks.
All our computations were done on a Connection Machine CM-2 using a conjugate
gradient linear equation solver (written in CMIS) which runs at $>5$~Gigaflops/sec.

The first exercise we had to undertake was to find a suitable value for $\kappa$
to give the desired pion mass. We ran at $\kappa=0.1680$, $0.1677$, $0.1675$,
and $0.1670$: our results indicate that $m_\pi<0.2/a$ for the first two values,
indicating that $\kappa_c$ is probably in this region. If we extrapolate
recently published results\cite{LANL} we find that
$\kappa_c\approx0.1676$, in agreement with our data. For $\kappa=0.1670$ we
find that $m_\pi\approx0.47/a$. We are currently running at $\kappa=0.1675$
which we hope will give us the desired value for $m_\pi$.

The major reason for the uncertainties in the preceding statements about
pion masses is that it is very hard to bring the system into equilibrium when
the pion mass is small because the relaxation and autocorrelation times of the
HMC algorithm are extremely large, at least $1,000$ units of MD time for the
lightest systems with $m_\pi<0.2/a$. 

We {\em should} expect this behavior.
The pion is the lightest excitation of QCD, because it is trying to be a
Goldstone boson. Unlike the quenched approximation where the lightest dynamical
excitation is a glueball, for full QCD the pion mass is the inverse correlation
length, $m_\pi=1/\xi$. Thus, for $m_\pi=0.2/a$ we expect that the correlation length
$\xi\approx5a$ in equilibrium; it is only reasonable to expect an algorithm to
take much longer to develop such long distance correlations than it takes to
equilibrate a quenched system with $\xi\approx a$. Theoretically, we expect the
characteristic relaxation time to be the same as the autocorrelation time, which
in turn is proportional to $\xi^z$ where $z$ is the dynamical critical exponent.
For HMC it is believed that $1\leq z\leq2$ \cite{BJP_ADK,SGupta_91a,SGupta_91b}.

How can we measure the correlation length of an ensemble of configurations? In
principle we could look at the asymptotic decay of some purely gluonic
correlation function, but in practice there would be no hope of extracting a
signal because of fluctuations in the disconnected part of such correlation
functions.

The obvious method of measuring $\xi$ is to measure the pion mass by measuring
the asymptotic decay of a pion correlation function $\langle\pi(0)
\pi(t)\rangle$. If we have a set of equilibrated configurations this is a
completely valid procedure, but consider what would happen if the Markov process
had not yet converged to the true equilibrium distribution. 
Because the system starts from
a configuration which does not have the true long distance correlations built in,
the unequilibrated configurations will tend to have $\xi<1/m_\pi$. Roughly
speaking we can say that they correspond more closely to the equilibrium
distribution of QCD with a smaller value of $\kappa$ -- we shall call
this the ``sea'' quark hopping parameter -- than the actual value
appearing in the action.
Of course, the actual distribution of some set of unequilibrated configurations
is some complicated mess, but it is certainly reasonable to assume that such a
shift in $\kappa$ is the dominant effect when near $\kappa_c$.

The hopping parameter does not only appear in the action, it is also explicitly
present in the pion operator $\pi$ used to measure the pion mass. Since this
hopping parameter has nothing to do with the dynamics of the system we shall
call it the ``valence'' hopping parameter. For our computations we expect that
the sea $\kappa$ approaches the valence $\kappa$ from below as the system
approaches equilibrium: in other words the correlation length estimated using
the correlation function for a ``valence'' pion will be larger than the actual
correlation length of the system until the system reaches equilibrium.

For a quenched computation or for an inexact dynamical quark
algorithm valence and sea parameters are different even in equilibrium.
One illustration of these ideas is that it is easy to measure a light pion mass
on a quenched configuration, even though the correlation length is much smaller
than the inverse pion mass.

The existence of a light valence pion is not evidence for a system containing
light dynamical pions unless the system can be shown to have
equilibrated to the correct probability distribution.

%
%
\begin{figure}
\epsfxsize=\columnwidth
\epsffile{figure1.ps}
\caption{
  Time history of the effective pion mass (distance $8$) for $\beta=5.30$ and
  $\kappa=0.1677$ on a $16^3\times 32$ lattice. Symbols are: ($\times$) HMC
  at $\delta\tau=0.0069$, ($+$) HMD at $\delta\tau=0.0069$, ($\bigcirc$) HMD
  at $\delta\tau=0.004$. Also shown is the blocksize$=40$ ($\Diamond$) HMC
  acceptance uncorrected for autocorrelations.
}
\label{HMD_pion} 
\end{figure}

In Figure~\ref{HMD_pion} we show the pion mass and acceptance rate as
a function of MD time
for our run at $\kappa=0.1677$. It is obvious that the system has not
equilibrated even after many hundreds of MD time units, so we can only deduce
the lower bound on the pion mass stated above. It is also clear that as the
correlation length grows the HMC acceptance rate falls, which requires reducing
the integration step size in order to keep a reasonable acceptance rate (the
portion of the HMC data shown was produced with a step size of
$\delta\tau=0.0069$ then switched to a step size of $\delta\tau=0.004$
at trajectory $711$.).
A preliminary study of the system indicates that this is primarily due to the
increasing effect of high frequency fluctuations in the fermionic contributions
to the action.

We decided it would be useful to see what results we would get if we used the HMD
algorithm instead of HMC for this system (i.e., if we omitted the Monte Carlo
accept/reject step). It is commonly accepted that this should introduce small
errors in physical quantities of order $\delta\tau^2$; to our surprise the HMD
algorithm with the same parameters gave completely wrong answers for the pion
mass. In fact other hints that the system contained light pions, such as the
large number of CG iterations required per step to reach a given residual, also
rapidly disappeared when running the HMD algorithm. When we reduced the
step size to $\delta\tau=0.004$, the HMD results were virtually identical. 
It would appear
that we would have to run at much smaller step sizes to reach the region were we
could even attempt to undertake a zero step-size extrapolation.

Our results so far indicate that there are no significant
problems for HMC computations in QCD with Wilson quarks, other than that they
are very expensive. On the other hand, we have not yet been able to carry out an
HMD run which gets even approximately the correct answer.

If we had carried out an HMD calculation with the same step sizes as
used in the present HMC computations,
we would have been forced to use a larger value of $\kappa$ in
order to measure a light pion mass. This mass would have been much smaller than
the inverse correlation length actually present in the configurations. Indeed,
we may well have been measuring valence observables on an almost-quenched
system, so we would have concluded that physics with light dynamical quarks was
very similar to quenched physics with the same masses. We would also have found
a much smaller relaxation time and autocorrelation time, and thus been misled
into thinking that full QCD calculation were much cheaper than they really are.

What are the implications of this for staggered quarks calculations
done using the HMD algorithm?
While our Wilson quark results do not tell us directly about the
behaviour of staggered quarks --- the effects we see might possibly just be
artefacts of Wilson quark dynamics --- they lead us to suggest that it is
necessary for a careful zero step-size extrapolation to be done for any HMD
calculation, with special attention required to verify that the system is
truly in equilibrium.

While we cannot directly compare the ``two flavour'' R-algorithm HMD staggered
quark results with any exact algorithm because no local action exists for such a
system, we hope to carry out a careful comparison of HMD and HMC with four light
flavors of  staggered quarks.

Lack of space prevents us from discussing our results on higher order
integration schemes.



\begin{thebibliography}{9}


\bibitem{LANL}
{\it QCD~with~Wilson Quarks II.}, Los Alamos Group,
LA-UR-91-528, UW DOE/ER/40614-4, COLO-HEP-254, June 1991.

\bibitem{BJP_ADK}
{\it Acceptances and Autocorrelations in Hybrid Monte Carlo},
A.D.~Kennedy and B.~Pendelton, Nuc. Phys. {\bf B20}(Proc. Suppl.) 118 (1991). 

\bibitem{SGupta_91a}
{\it Dynamic Critical Properties of Hybrid Monte Carlo},
S. Gupta, CERN-TH.~6178/91. 

\bibitem{SGupta_91b}
{\it Acceptance and Autocorrelation in Hybrid Monte Carlo},
S. Gupta, CERN-TH.~6109/91. 

\bibitem{Campostrini_90}
M.~Campostrini and P.~Rossi, Nuc. Phys. {\bf B329} 753 (1990). 

\bibitem{Creutz_89}
M.~Creutz and A.~Gocksch, Phys. Rev. Let. {\bf 63}, 9 (1989). 

\end{thebibliography}
\end{document}